\documentclass[conference]{IEEEtran}
\IEEEoverridecommandlockouts
\usepackage{cite}
\usepackage{amsmath,amssymb,amsfonts}
\usepackage{algorithmic}
\usepackage{graphicx}
\usepackage{textcomp}
\usepackage{xcolor}

\usepackage{booktabs}
\usepackage[ruled,linesnumbered]{algorithm2e}
\usepackage{makecell}
\usepackage{soul}
\usepackage[a4paper, total={184mm,239mm}]{geometry}
\def\BibTeX{{\rm B\kern-.05em{\sc i\kern-.025em b}\kern-.08em
    T\kern-.1667em\lower.7ex\hbox{E}\kern-.125emX}}
\begin{document}

\title{LaMoS: Enabling Efficient \underline{La}rge Number \underline{Mo}dular Multiplication through \underline{S}RAM-based CiM Acceleration
}

\author{
\IEEEauthorblockN{Haomin Li$^{1}$, Fangxin Liu$^{1,2,*}$, Chenyang Guan$^{1}$, Zongwu Wang$^{1,2}$, Li Jiang$^{1,2,*}$, Haibing Guan$^{1}$}
\IEEEauthorblockA{1. School of Computer Science, Shanghai Jiao Tong University,\,\,\,2. Shanghai Qi Zhi Institute}
\IEEEauthorblockA {*Corresponding Author \,\,\,\,\,\,\,\ \{haominli,liufangxin, ljiang\_cs\}@sjtu.edu.cn
}
\thanks{
This work is supported by the National Key Research and Development Program of China (2024YFE0204300), the National Natural Science Foundation of China (Grant No.62402311), and Natural Science Foundation of Shanghai (Grant No.24ZR1433700). Fangxin Liu and Li Jiang are the corresponding authors.}
}

\maketitle

\begin{abstract}
Barrett’s algorithm is one of the most widely used methods for performing modular multiplication, a critical nonlinear operation in modern privacy computing techniques such as homomorphic encryption (HE) and zero-knowledge proofs (ZKP). Since modular multiplication dominates the processing time in these applications, computational complexity and memory limitations significantly impact performance. 
Computing-in-Memory (CiM) is a promising approach to tackle this problem.
However, existing schemes currently suffer from two main problems: 1) Most works focus on low bit-width modular multiplication, which is inadequate for mainstream cryptographic algorithms such as elliptic curve cryptography (ECC) and the RSA algorithm, both of which require high bit-width operations; 2) Recent efforts targeting large number modular multiplication rely on inefficient in-memory logic operations, resulting in high scaling costs for larger bit-widths and increased latency.
To address these issues, we propose LaMoS, an efficient SRAM-based CiM design for large-number modular multiplication, offering high scalability and area efficiency. First, we analyze the Barrett’s modular multiplication method and map the workload onto SRAM CiM macros for high bit-width cases. Additionally, we develop an efficient CiM architecture and dataflow to optimize large-number modular multiplication. Finally, we refine the mapping scheme for better scalability in high bit-width scenarios using workload grouping. Experimental results show that LaMoS achieves a $7.02\times$ speedup and reduces high bit-width scaling costs compared to existing SRAM-based CiM designs.
\end{abstract}


\section{Introduction}

Privacy computing, a key approach to ensuring data security, has gained significant attention in recent years due to rising concerns over privacy and personal data protection on the Internet~\cite{xiao2012security,li2020review,zhang2021survey,liu2023hyperattack,li2024hyperfeel,li2025attack}. It encompasses a variety of applications, such as homomorphic encryption (HE)~\cite{acar2018survey}, which safeguards the privacy of user data and models, and zero-knowledge proofs (ZKP)~\cite{goldwasser2019knowledge,fiege1987zero}, an emerging cryptographic protocol that allows a verifier to confirm the truth of a statement without revealing any additional information. These applications rely on public key cryptography algorithms like elliptic curve cryptography (ECC)~\cite{koblitz1987elliptic} and RSA~\cite{rivest1978method}, where modular multiplication plays a critical role in preventing overflow.

Notably, modular multiplication in these algorithms requires extremely high bit-width implementations—ECC demands at least 224 bits~\cite{chen2023digital}, while RSA typically requires 1024 bits or more~\cite{bos2009security}. In most cryptographic schemes, higher bit-width operations correspond to greater security~\cite{schneier2007applied}. However, this high bit-width processing poses significant challenges for efficient hardware implementation: 1) Efficient algorithms like Barrett~\cite{barrett1986implementing} and Montgomery~\cite{montgomery1985modular} rely on fast multipliers, and implementing them for high bit-widths increases latency and area complexity. 2) Modular multiplication generates numerous intermediate results, placing heavy demands on storage and memory bandwidth. 

\begin{figure}[tp]
    \setlength{\abovecaptionskip}{1pt}
    \setlength{\belowcaptionskip}{1pt}
    \centering
    \includegraphics[width=1\linewidth]{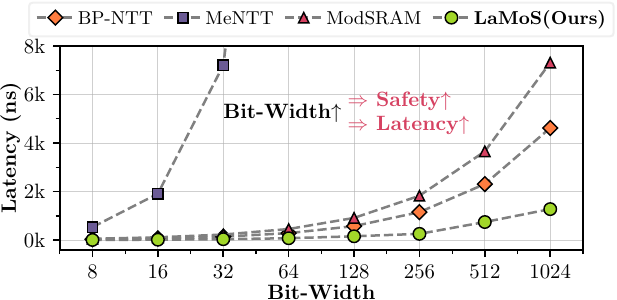}
    \caption{Performance Comparison with previous works over various bit-widths.}
    \label{exp-fig: intro}
\end{figure}
Previous works on FPGA-based modular multiplication have focused on optimizing the use of DSPs and LUTs to enhance computational efficiency~\cite{langhammer2021efficient,zhang2021pipezk,opasatian2023lookup,devlin2019blockchain,ozturk2020design,gribok2024fpga,liu2025allmod}. However, these approaches overlook the significant data handling overhead characteristic of real-world privacy computing applications. For instance, PipeZK~\cite{zhang2021pipezk} requires nearly 3 TB/s bandwidth at 100 MHz to compute a ZKP scheme, which is impractical for current systems. To address this issue, computing-in-memory (CiM)~\cite{mutlu2022modern, sebastian2020memory,liu2025asdr,wang2024compass,liu2024paap} has emerged as a promising platform to reduce data handling overhead. It has been widely explored for accelerating privacy computing applications, including advanced encryption standard (AES) and homomorphic encryption~\cite{reis2022imcrypto,ku2024modsram,li2022mentt,zhang2023bp,park2022rm,nejatollahi2020cryptopim,li2023accelerating}.

However, these acceleration designs primarily target cryptographic algorithms with low bit-width requirements, like post-quantum cryptography (PQC)~\cite{bernstein2017post}, which typically only demand bit-widths of 64 or less, allowing modular multiplication to be implemented with minimal overhead. In contrast, many privacy computing applications rely on cryptographic algorithms with much higher bit-width requirements, such as ZKP based on ECC, which requires several hundred bits for modular multiplication.

As illustrated in Figure~\ref{exp-fig: intro}, we compare the performance of different works~\cite{ku2024modsram,li2022mentt,zhang2023bp} across various bit-widths\footnote{These works do not report performance across all bit-widths, so we scaled the results according to the complexity of the designs.}. It is clear that extending existing designs naively to higher bit-widths introduces significant latency overhead, as these schemes are not optimized for large-number modular multiplication. The latest work, ModSRAM~\cite{ku2024modsram}, focuses on accelerating large number modular multiplication for ECC. However, the performance of the proposed accelerator when integrated into SRAM in-memory logic is inadequate in two key areas:
\begin{itemize}
    \item \textbf{High latency and resource inefficiency:} For 256-bit modular multiplication, ModSRAM requires 767 cycles at 400 MHz, which is insufficient for privacy computing applications. This high latency also negatively impacts the CiM design, as it locks the SRAM array for too long. For a 256-bit implementation, the 6KB ($64\times 256$) SRAM-based CiM array, where operands are stored, remains locked for 767 cycles, restricting its use solely to modular multiplication. In other words, high latency amplifies resource waste in CiM arrays.

    \item \textbf{High scaling cost:} Since the implementation relies on parallel bit-wise logic operations in SRAM, scaling ModSRAM to higher bit-widths is extremely costly. For example, increasing from 256-bit to 512-bit would require more than 3,000 cycles to complete a modular multiplication, unless additional SRAM arrays are introduced to maintain higher parallelism in the bit-wise logic.

\end{itemize}
To address the high latency and scaling challenges in large-number modular multiplication, we propose \textbf{LaMoS}, an efficient SRAM-based CiM design. First, we analyze the key operations in Barrett's modular multiplication algorithm and develop a mapping scheme that divides the computation of large number multiplications into smaller workloads, which are mapped onto the SRAM-based CiM macro. Then, we design an architecture and dataflow that utilizes parallelism across CiM macros to achieve low-latency large-number modular multiplication. Next, we introduce a workload grouping-based mapping optimization to improve performance when scaling to higher bit-widths by reducing redundant CiM macro usage.

Overall, our work has the following contributions:
\begin{itemize}
    \item We design LaMoS, a high-efficiency SRAM-based CiM architecture for large-number modular multiplication. By exploiting workload parallelism across multiple CiM macros, LaMoS significantly reduces latency for large-number modular multiplication.
    \item We propose a novel mapping strategy for large-number multiplication within SRAM-based CiM macros. By partitioning the large-number multiplication into smaller, manageable workloads and mapping them efficiently onto the macros, we enable high-performance parallel computation, which forms the basis for an optimized Barrett's modular multiplication.
    \item We optimize the workload mapping process to minimize latency as bit-widths increase. Through grouping workloads and eliminating redundant computations, LaMoS can maximize CiM macro utilization. Evaluations demonstrate that LaMoS delivers high area efficiency, achieving a $6\times$ speedup over ModSRAM and significantly reducing scaling costs for 1024-bit and 2048-bit operations. 
\end{itemize}

\section{Backgrounds}

\begin{algorithm}[tb]
\caption{Barrett's Modular Multiplication~\cite{barrett1986implementing}.}
\label{alg:barrett}


\let\oldnl\nl
\newcommand{\nonl}{\renewcommand{\nl}{\let\nl\oldnl}}


\KwData{
three $n$-bit unsigned numbers $A$, $B$, and $M$
}
\KwResult{$A\times B\ mod\ M$}

\colorbox[RGB]{255, 255, 255}{
    \makebox[0.9\linewidth]{
        \hspace{-0.47cm}
        \textbf{Offline:}\hfill
    }
}

\colorbox[RGB]{242, 242, 242}{
    \makebox[0.9\linewidth]{
        \hspace{-0.00cm}
        $M' \leftarrow \lfloor \frac{2^{2n}}{M} \rfloor$ \hfill \textbf{Precompute}
    }
}

\colorbox[RGB]{255, 255, 255}{
    \makebox[0.9\linewidth]{
        \hspace{-0.47cm}
        \textbf{Online:} \hfill
    }
}

\colorbox[RGB]{251, 229, 214}{
    \makebox[0.9\linewidth]{
        \hspace{-0.00cm}
        $C\leftarrow A \times B$ \hfill \textbf{Multiplier}
    }
}

\colorbox[RGB]{251, 229, 214}{
    \makebox[0.9\linewidth]{
        \hspace{-0.00cm}
        $u \leftarrow \lfloor \frac{C}{2^{n-1}}\rfloor\times M'$ \hfill \textbf{Multiplier}
    }
}

\colorbox[RGB]{222, 235, 247}{
    \makebox[0.9\linewidth]{
        \hspace{-0.00cm}
        $E\leftarrow \lfloor \frac{u}{2^{n+1}}\rfloor$ \hfill \textbf{Shift}
    }
}

\colorbox[RGB]{251, 229, 214}{
    \makebox[0.9\linewidth]{
        \hspace{-0.00cm}
        $P\leftarrow E\times M$ \hfill \textbf{Multiplier}
    }
}

\colorbox[RGB]{226, 240, 217}{
    \makebox[0.9\linewidth]{
        \hspace{-0.00cm}
        $T \leftarrow C-P$ \hfill \textbf{Subtractor}
    }
}

\colorbox[RGB]{226, 240, 217}{
    \makebox[0.9\linewidth]{
        \hspace{-0.00cm}
        $R \leftarrow T\ or\ T-M\ or\ T-2M$\ \hfill \textbf{Subtractors}
    }
}
\end{algorithm}

\subsection{Modular Multiplication}

Let $A$, $B$ and $M$ be $n$-bit unsigned inputs, where $C = AB$ is a $2n$-bit unsigned value. Modular multiplication computes the remainder $R$ of $C$ divided by $M$:
\begin{equation}
    \begin{aligned}
    R = AB\ mod\ M = C\ mod\ M
    \end{aligned}
\end{equation}

Barrett’s algorithm is one of the most commonly used methods for modular multiplication, as it replaces the slow and costly division operation with several multiplications. The detailed steps of Barrett's algorithm are outlined in Algorithm~\ref{alg:barrett}.
First, $C = A\times B$ is computed by an $n$-bit multiplier. 
Then, the quotient $E$ is estimated by multiplying the lower low $n+1$ bits of $C$ by $M'$, where $M'$ is the precomputed reciprocal of modulo $M$, and then truncating the result to $n+1$ bits.
Since the modulo $M$ is often static in a privacy computing application, $M'$ can be computed offline as $\lfloor \frac{2^{2n}}{M} \rfloor$.
After that, $EM$ is subtracted from $C$ to obtain an intermediate value $T$.
Finally, $T$ is subtracted by $M$ until the final result lies within the range $[0,M)$. This final reduction step typically requires at most two subtractions, which can be efficiently implemented using cascaded subtractors and a multiplexer.


\begin{figure*}[thbp]
    \setlength{\abovecaptionskip}{1pt}
    \setlength{\belowcaptionskip}{1pt}
    \centering
    \includegraphics[width=1\linewidth]{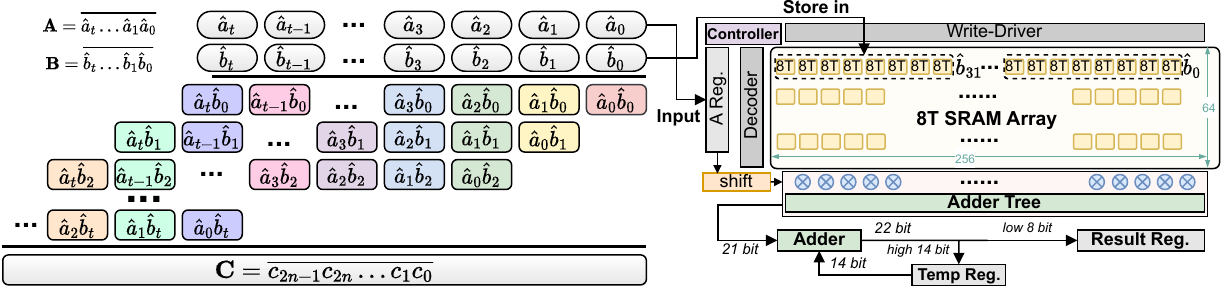}
    \caption{Large Number Multiplication Workload Mapping onto the SRAM Macro.}
    \label{fig: framework}
\end{figure*}

\subsection{Cryptographic CiM}
Several cryptographic CiM accelerators have been developed based on both SRAM~\cite{li2022mentt,zhang2023bp} and ReRAM array~\cite{park2022rm,nejatollahi2020cryptopim,li2023accelerating}, with a focus on reducing the heavy data movement overhead commonly found in privacy computing applications. These accelerators typically target cryptographic schemes with lower bit-width requirements, such as AES and PQC. They often optimize at the algorithmic level, focusing on techniques like the number theoretic transform (NTT), or at the application level, for example, in homomorphic encryption. ModSRAM~\cite{ku2024modsram} is a more recent design that addresses large-number modular multiplication in ECC, which demands higher bit-width representations ranging from 224 to 512 bits. ModSRAM utilizes SRAM in-memory logic to implement modular multiplication based on a modified interleaved algorithm. However, its performance remains suboptimal. For an $n$-bit modular multiplication, ModSRAM requires $3n+1$ cycles to complete the calculation, which falls short of the requirements for privacy computing applications.





\subsection{Motivation}
Recent CiM architectures primarily focus on cryptographic schemes such as AES and PQC, which represent only a small subset of cryptographic applications. Many widely used schemes like ECC and RSA require extremely high bit-width implementations, demanding support for large-number arithmetic operations. Additionally, in most cryptographic schemes, increasing the bit-width enlarges the key space, making the system harder to crack and thereby enhancing security.

Thus, accelerating high bit-width arithmetic operations, particularly large-number modular multiplication, is crucial for cryptography. However, existing CiM designs generally target low bit-width multiplication, or they are difficult to optimize for larger bit-widths, leading to significant area or latency overhead as the bit-width scales. While SRAM-based CiM designs implement bit-wise in-memory logic for large-number modular multiplication, like ModSRAM, they suffer from inefficiency and high scaling costs.

In summary, a more efficient solution with lower scaling costs for large-number modular multiplication is still missing in above works, which we believe will be the trend for future privacy computing accelerators. SRAM-based CiM macros with multiply-and-accumulate (MAC) capabilities~\cite{zhu2022comb} offer greater computational power, potentially leading to reduced latency and scaling costs compared to SRAM-based CiM designs using logic operations. Furthermore, Barrett's algorithm, which relies on multiplication, has a lower latency overhead compared to interleaved algorithms. Therefore, we propose an efficient and scalable modular multiplication architecture based on Barrett's algorithm, harnessing the MAC capabilities of SRAM CiM macros to achieve superior performance.

\begin{figure}[bp]
    \setlength{\abovecaptionskip}{1pt}
    \setlength{\belowcaptionskip}{1pt}
    \centering
    \includegraphics[width=1\linewidth]{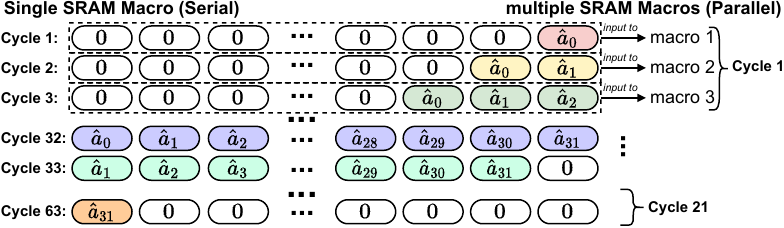}
    \caption{Input Timing Flow for 256-bit multiplication with single/multiple SRAM macro(s). Number of SRAM macros is set to 3 in the example.}
    \label{fig: input flow}
\end{figure}

\section{Proposed Design} 



\subsection{Workload Segmentation and Mapping}

To address the computational demands of Barrett's algorithm, which involves three large-number multiplications (\textcolor{orange}{orange part} in  Algorithm~\ref{alg:barrett}), we propose a systematic approach for mapping these operations onto SRAM-based CiM MAC macros. Our design leverages the $64 \times 256$ 8T SRAM macro, known for its capability to execute 32 8-bit MAC operations per cycle. We decompose the large-number multiplication into several smaller workloads, each comprising 32 8-bit multiplication operations. These workloads are efficiently mapped onto the SRAM macro, allowing the large-number multiplication to be executed with minimal latency and using a few low bit-width accumulators. 

Specifically, we perform the following transformation on the large numbers $A$ and $B$, involved in the multiplication, by expressing them as a weighted sum of 8-bit numbers:
\begin{equation}
    \begin{aligned}
        A &= \sum_{i=0}^{n-1} (2^{i}a_{i})
        = \sum_{i=0}^{\frac{n}{8}-1} \sum_{j=0}^{7}(2^{8i+j}a_{8i+j})
        = \sum_{i=0}^{\frac{n}{8}-1} \hat{a}_i\times 2^{8i}\\
        B &= \sum_{i=0}^{n-1} (2^{i}b_{i})
        = \sum_{i=0}^{\frac{n}{8}-1} \sum_{j=0}^{7}(2^{8i+j}b_{8i+j})
        = \sum_{i=0}^{\frac{n}{8}-1} \hat{b}_i\times 2^{8i}\\
    \end{aligned}
\end{equation}
where $\hat{a}_i = \sum_{j=0}^{7}(2^{j}a_{8i+j})$ and $\hat{b}_i = \sum_{j=0}^{7}(2^{j}b_{8i+j})$, both of which are 8-bit numbers.
Next, the multiplication can be represented as a weighted sum of the products of these 8-bit multiplications:
\begin{equation}
    \begin{aligned}
        C &= A \times B
        = \sum_{i=0}^{\frac{n}{8}-1}\sum_{j=0}^{\frac{n}{8}-1} \hat{a}_i \times 2^{8i} \times \hat{b}_j \times 2^{8j} \\
        &= \sum_{i=0}^{\frac{n}{8}-1}\sum_{j=0}^{\frac{n}{8}-1} \hat{a}_i \times \hat{b}_j \times 2^{8(i+j)} \\
    \end{aligned}
\end{equation}

\begin{figure}[tp]
    \setlength{\abovecaptionskip}{1pt}
    \setlength{\belowcaptionskip}{1pt}
    \centering
    \includegraphics[width=1\linewidth]{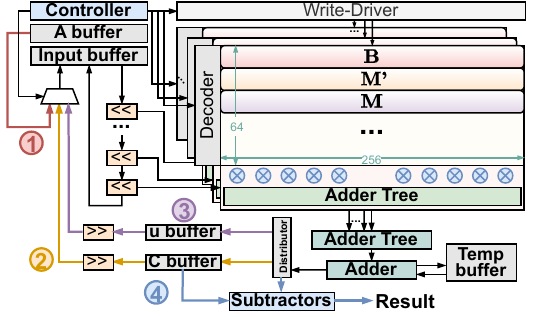}
    \caption{LaMoS Architecture and Dataflow for efficient large number modular multiplication.}
    \label{fig: arch}
    \vspace{-0.1cm}
\end{figure}

\begin{figure*}[tp]
    \setlength{\abovecaptionskip}{1pt}
    \setlength{\belowcaptionskip}{1pt}
    \centering
    \includegraphics[width=1\linewidth]{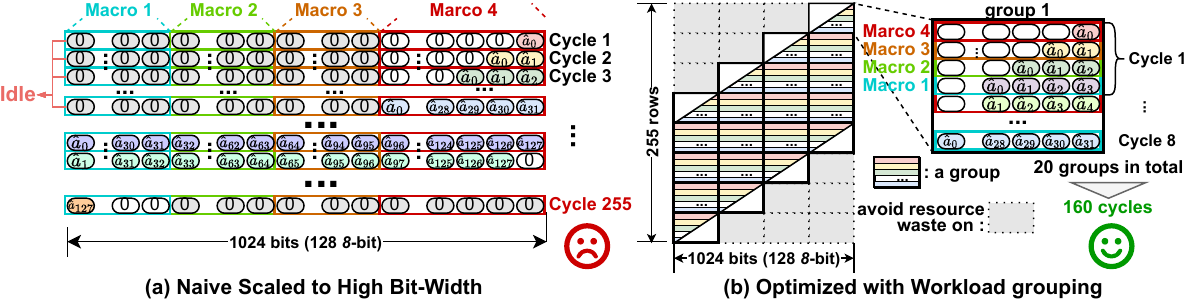}
    \caption{LaMoS Execution for high bit-width modular multiplication. (a) Naively Scaling. (b) Optimization with workload grouping.}
    \label{fig: highbit}
    \vspace{-0.1cm}
\end{figure*}

Figure~\ref{fig: framework} illustrates the vertical structure of the multiplication between $A$ and $B$, where each entry in the vertical layout corresponds to the product of two 8-bit numbers $\hat{a}_i$ and $\hat{b}_j$.
Each multiplication $\hat{a}_i\hat{b}_j$ can be regarded as a separate workload. 
Therefore, a multiplication of two $n$-bit numbers can be viewed as the accumulation of $t^2$ workloads, where $t = \lceil \frac{n}{8} \rceil$. 
Additionally, within the vertical equation, workloads in the same column share the same weighting.
For example, the results of $\hat{a}_2\hat{b}_0$, $\hat{a}_1\hat{b}_1$, $\hat{a}_0\hat{b}_2$ all have the same weighting factor of $2^{16}$. This allows us to map the workloads in each column to a single MAC operation within the CiM macro, thereby optimizing the overall computation.

Figure~\ref{fig: framework} also illustrates the mapping process for a $256$-bit multiplication. 
We begin by mapping the operand $B$ onto a row of SRAM as $32$ $8$-bit numbers. During each cycle, to perform a single MAC operation of large-number multiplication, we provide the necessary inputs from $A$ for one column of workloads.

The input stream configuration for a single MAC macro is detailed in Figure~\ref{fig: input flow}.
In the first cycle, the inputs consist of $31$ $8'b0$s and $\hat{a}_0$. 
In the second cycle, the inputs are 30 $8'b0$, $\hat{a}_{1}$, and $\hat{a}_{0}$. Such a workload mapping ensures that the SRAM-based CiM MAC macro is utilized efficiently for workload computation. For a $256$-bit multiplication, this approach allows all workloads to be completed within $63$ cycles on a single MAC macro.


The peripheral accumulation circuit for the MAC results\footnote{The MAC result at each cycle is a $8 + 8 + \log_2(32) = 21$-bit value.} is detailed in Figure~\ref{fig: framework}. Each cycle generates a $21$-bit MAC result, where the low-order $8$ bits of this $21$-bit value directly contribute to the final result. This is because the weighting difference between MAC results of adjacent cycles is $2^8$, making the low-order $8$ bits of each cycle's result the significant part of the final result.

During each cycle, the $21$-bit MAC result is split into two parts: the low-order $8$ bits and the high-order $13$ bits. The low-order $8$ bits are added to the final result register. To handle the high-order bits, the high-order $13$ bits of the current cycle's result are added to the high-order $14$ bits of the $22$-bit addition result (with $1$ carry bit) from the previous cycle using a $21$-bit adder. The result of this addition provides the high-order part for the current cycle, with its low-order $8$ bits stored in the final result register and the high-order $14$ bits kept in a temporary register to be carried over and added in the next cycle. This process ensures the correct accumulation of results across all cycles.



\subsection{Architecture Design}
Based on the proposed mapping scheme, we design LaMoS, an efficient and scalable architecture for CiM-based large-number modular multiplication, as shown in Figure~\ref{fig: arch}.

For single large-number multiplications, LaMoS uses a scalable parallel design with multiple CiM macros. To perform modular multiplication, we incorporate peripheral circuits and dataflows that optimize computation efficiency. We use a parallel acceleration approach with multiple macros. Multiple columns of the workload, as illustrated in Figure~\ref{fig: input flow}, are distributed across these macros. The value of $B$ is replicated and stored in each macro. In each cycle, each macro processes one column of the workload independently.
For example, in a 256-bit multiplication with three macros, during the first cycle, macro \#1 processes $31$ $8'b0$ and $\hat{a}_0$, macro \#2 processes $30$ $8'b0$ along with $\hat{a}_1$ and $\hat{a}_0$, and macro \#3 processes $29$ $8'b0$ along with $\hat{a}_2$, $\hat{a}_1$, and $\hat{a}_0$.

The peripheral circuits designed to support large-number modular multiplication are illustrated in Figure~\ref{fig: arch}.
LaMoS includes a low-bit-width adder and temporary registers to accumulate computation results from each cycle. An adder tree aggregates results from parallel macros, with minimal overhead given that a small number of macros (between 2 and 8) is sufficient for optimal latency. For input data handling, we use a shift array to provide data with different shifts to the appropriate macro each cycle. We incorporate buffers, a multiplexer (MUX), and a distributor to manage the three different multiplications.
\begin{itemize}
    \item First multiplication ($A \times B$): The MUX directs $A$ to the input buffer for computation. The resulting product $C$ is stored in the C buffer via the distributor.
    \item Second multiplication ($\lfloor \frac{C}{2^{n-1}} \rfloor \times M'$): The value of $C$ from the C buffer is shifted and sent to the input buffer. The result $u$ is stored in the u buffer.
    \item Third multiplication ($E \times M$): The value $u$ from the u buffer is shifted to obtain $E$, which is then sent to the input buffer. Finally, the distributor routes the result $P$ to a subtractor, where $P$ is subtracted from $C$ to obtain $T$. The subtractor then performs additional subtractions of $M$ from $T$ to complete the refinement process.
\end{itemize}

\subsection{Mapping Optimization for Scaling to Higher Bit-Width}
As LaMoS is scaled to handle higher bit-widths (e.g., beyond 256 bits), the latency of computations increases significantly. Figure~\ref{fig: highbit}(a) demonstrates the input stream when mapping high bit-width multiplications onto the architecture. For instance, in a 1024-bit multiplication with LaMoS configured with 4 macros, the 1024-bit number is expressed as a weighted sum of 128 $8$-bit numbers. Since each $64 \times 256$ macro supports MAC operations on only 32 $8$-bit inputs, the input stream must be divided into 4 slices for parallel processing by the 4 macros. Consequently, 255 cycles are required to complete the multiplication.

To enhance performance, we analyze the workload distribution and observe that some macros are often idle. For example, in the first 92 rows of the input stream, macro 1 consistently processes $0$s. This idleness results from varying workloads per column in Figure~\ref{fig: framework}, leading to excessive padding with $0$s. Figure~\ref{fig: highbit}(b) shows a $255$-row by $1024$-bit input stream for a 1024-bit multiplication, highlighting large amounts of padded $0$s (gray and white regions). The \textcolor{gray}{gray} regions correspond to cycles where all inputs to a macro are $0$s, causing the macros to be idle.

To this end, we propose a grouping-based mapping optimization. We divide the input streams into groups of $32$-row by $128$-bit each, ensuring that a single macro processes exactly one row per group. Redundant groups (\textcolor{gray}{gray} region) with only $0$s are discarded. Non-redundant groups are processed by multiple macros in parallel, requiring only 8 cycles per group, given 4 macros. This optimization reduces the workload by $37.5\%$ for 1024-bit multiplications and eliminates macro idleness, achieving a final latency of 160 cycles.

\section{Evaluations}

\subsection{Evaluation Methodology}

\textbf{Implementation and Simulation}: 
We implemented LaMoS using Verilog and synthesized the RTL with Synopsys Design Compiler~\cite{2019DC} at 400MHz, targeting the TSMC 28nm standard library to evaluate the area. For the SRAM-based CiM macro configuration, we referred to the data reported in \cite{zhu2022comb}. To assess latency, we developed a cycle-accurate simulator that models the micro-architecture behavior across various bit-widths. Balancing area and latency, we evaluated LaMoS configured with $2$ $64\times 256$ CiM macros.


\begin{figure}[bp]
    \vspace{-0.1cm}
    \setlength{\abovecaptionskip}{1pt}
    \setlength{\belowcaptionskip}{1pt}
    \centering
    \includegraphics[width=0.95\linewidth]{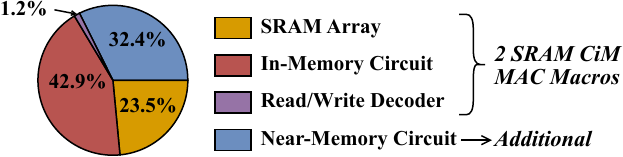}
    \caption{LaMoS Area Breakdown.}
    \label{exp-fig: breakdown}
\end{figure}

\textbf{Baselines}: 
We compare the LaMoS with the following designs:
(1) SRAM-based CiM designs: MeNTT~\cite{li2022mentt} and BP-NTT~\cite{zhang2023bp}; (2) ReRAM-based CiM designs: RM-NTT~\cite{park2022rm}, CryptoPIM~\cite{nejatollahi2020cryptopim}, and X-Poly~\cite{li2023accelerating}; (3) ModSRAM~\cite{ku2024modsram},  our primary baseline, as it is the only SRAM-based CiM design focused on large-number modular multiplication. Since these designs do not specifically target large number modular multiplication and only report performance on low bit-width operations, we scale their results to high bit-widths based on the theoretical complexity of each design\footnote{BP-NTT utilizes Montgomery's modular multiplication, so we account for the additional overhead of the Montgomery transformation.}.


\subsection{Performance}

\begin{table*}[tp]
\centering
\setlength{\abovecaptionskip}{1pt}
\setlength{\belowcaptionskip}{1pt}
\caption{Comparison on modular multiplication in PIM designs.}
\label{tab: overall}
\resizebox{\linewidth}{!}{
\begin{tabular}{cccccccc}
\Xhline{1.5px}
Reference & \textbf{LaMoS (ours)} & ModSRAM~\cite{ku2024modsram} & MeNTT~\cite{li2022mentt} & BP-NTT~\cite{zhang2023bp} & RM-NTT~\cite{park2022rm} & CryptoPIM~\cite{nejatollahi2020cryptopim} & X-Poly~\cite{li2023accelerating} \\ \Xhline{1.5px}
Application Type & ECC\ \& RSA  & ECC & PQC NTT & PQC NTT & HE NTT & PQC NTT & PQC NTT\\ \hline
Computation Method & Barrett & Direct & Direct & Montgomery & Montgomery & Montgomery/Barrett & Barrett \\ \hline
technology & 28nm & 65nm & 45nm & 28nm & 28nm & 45nm & 45nm\\ \hline
Cell Type & 8T SRAM & 8T SRAM & 6T SRAM & 6T SRAM & ReRAM & ReRAM & ReRAM \\ \hline
Array size & 2$\times$64$\times$256 & 64$\times$256 & 4$\times$162$\times$256 & 4$\times$256$\times$256 & 64$\times$4$\times$128$\times$128 & 512$\times$512 & 16$\times$128$\times$128 \\ \hline
Frequency (MHz) & 400 & 420 & 151 & 3.8k & 400 & 909 & 400 \\ \hline
Bitwidth & \textbf{Any Bit-width} & 256 & 14/16/32 & 2/4/8/16/32/64 & 14/16 & 16/32 & 16 \\ \hline
Cycles & \textbf{104} & 767 & 66049 & 4395 & - & - & - \\ \hline 
Area ($mm^2$) & \textbf{0.11} & 0.053 & 0.36 & 0.063 & - & 0.152 & 0.27 \\ \Xhline{1.5px}
Latency$\times$Area ($\mu s\times mm^2$) & \textbf{0.029} & 0.09 & 157.47 & \makecell{0.073} & - & - & -\\ \Xhline{1.5px}
\end{tabular}
}
\vspace{-0.3cm}
\end{table*}


Table~\ref{tab: overall} compares LaMoS with the baseline architectures. The cycle counts of the baselines are scaled to a 256-bit configuration for fair comparison. To evaluate overall performance, we use the $Latency \times Area$ metric, where lower values indicate better performance. Unlike existing SRAM-based CiM architectures, LaMoS is based on Barrett's algorithm, which results in extremely low latency, requiring only $104$ cycles for a $256$-bit modular multiplication. With the proposed mapping optimization, LaMoS can efficiently scale to support arbitrary bit-width computations.
In terms of overall performance, LaMoS demonstrates a $3\times$ improvement over the latest architecture, ModSRAM. Additionally, compared to BP-NTT, which uses the efficient Montgomery algorithm for modular multiplication, LaMoS achieves $2.5\times$ higher area efficiency by employing a more specialized architecture and dataflow that fully leverages the computational power of CiM macros.

For ReRAM-based architectures, directly scaling modular multiplication to higher bit-widths on crossbars leads to significant area overheads, both in terms of the number of crossbars and the required ADCs. It is important to note that our proposed workload mapping and optimization techniques are also well-suited for ReRAM-based CiM macros. As such, designing efficient ReRAM-based architectures for modular multiplication will be a key focus of our future research.

Figure~\ref{exp-fig: breakdown} provides the area breakdown of LaMoS, where the SRAM array occupies the largest portion. The near-memory circuits account for less than 35\% of the total area, demonstrating that our design incurs minimal overhead when extending from the CiM macros. Furthermore, the simplicity of the additional circuits highlights that LaMoS can be practically implemented with ease.

\subsection{Ablation Study}

\begin{figure}[tp]
    \setlength{\abovecaptionskip}{1pt}
    \setlength{\belowcaptionskip}{1pt}
    \centering
    \includegraphics[width=0.95\linewidth]{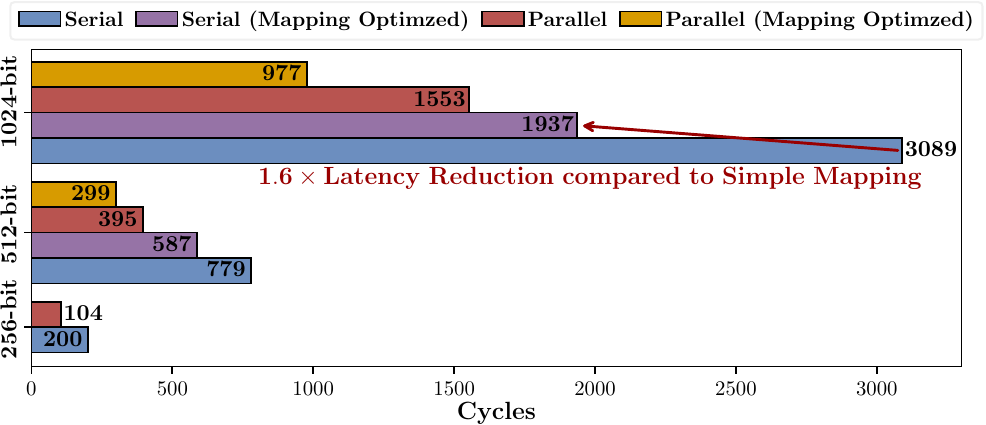}
    \caption{Detailed analysis of contributions.}
    \label{exp-fig: ablation}
    \vspace{-0.2cm}
\end{figure}

Figure~\ref{exp-fig: ablation} presents the results of the ablation study for LaMoS. The ``Serial'' configuration uses a single CiM macro, while the ``Parallel'' configuration employs two CiM macros. Mapping optimization is applied when the bit-width exceeds $256$ bits. For all bit-widths, the parallel LaMoS design achieves at least a $1.9\times$ speedup compared to the serial LaMoS. This demonstrates the effectiveness of parallel acceleration in our architecture, which is further explored in Section~\ref{subsection: DSE}. For bit-widths greater than 256 bits, LaMoS with mapping optimization achieves a speedup of $1.32\times$ and $1.59\times$ over the non-optimized version. This improvement is attributed to the workload grouping strategy, which minimizes idle time in the CiM macros and enhances the overall execution efficiency of LaMoS.



\subsection{Scalabitily Analysis}

\begin{figure}[bp]
    \vspace{-0.2cm}
    \setlength{\abovecaptionskip}{1pt}
    \setlength{\belowcaptionskip}{1pt}
    \centering
    \includegraphics[width=1\linewidth]{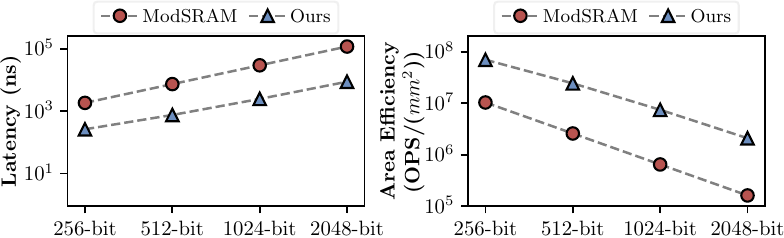}
    \caption{Comparison with ModSRAM\cite{ku2024modsram} on various bit-width.}
    \label{exp-fig: scale}
\end{figure}

To evaluate the scalability of LaMoS, we compare its latency and area efficiency with that of ModSRAM across different bit-widths, as shown in Figure~\ref{exp-fig: scale}. We use the same macro configuration as detailed in Table~\ref{tab: overall} to ensure a consistent basis for performance comparison at higher bit-widths.
For a 2048-bit width, ModSRAM experiences a latency exceeding $110,000$ ns and an area efficiency of less than $200,000$ OPS/mm$^2$. This performance degradation is primarily due to: (1) the triple-linear theoretical complexity of ModSRAM’s bit-logic-based modular multiplication algorithm with increasing bit-width, and (2) the need for multiple cycles to perform bit logic operations at high bit-widths, which can be completed in a single cycle at lower bit-widths, further amplifying latency issues. In contrast, LaMoS maintains a latency of less than $9,000$ ns and an area efficiency greater than $2,000,000$ OPS/mm$^2$ at the $2,048$ bits. These improvements are attributed to LaMoS’s efficient workload mapping and optimization techniques, as well as its multi-macro parallel design.

\subsection{Design Space Exploration}
\label{subsection: DSE}

\begin{figure}[tp]
    \setlength{\abovecaptionskip}{1pt}
    \setlength{\belowcaptionskip}{1pt}
    \centering
    \includegraphics[width=0.94\linewidth]{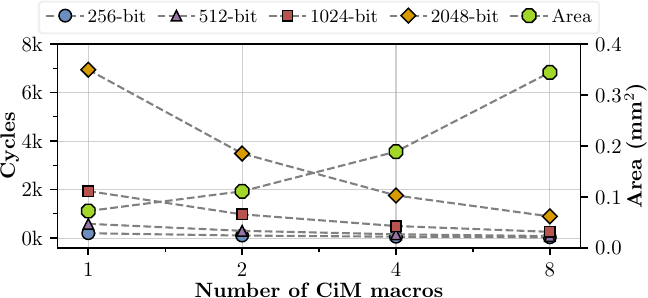}
    \caption{Design Space Exploration.}
    \label{exp-fig: design space}
    \vspace{-0.2cm}
\end{figure}
We conduct a design space exploration for LaMoS to assess its scalability and performance across various configurations. Figure~\ref{exp-fig: design space} illustrates the area results for LaMoS with different numbers of CiM macros and the corresponding performance at varying bit-widths.
The area of LaMoS increases linearly with the number of CiM macros, demonstrating minimal overhead for scaling up the architecture to support higher parallelism. With $4$ CiM macros, LaMoS achieves fewer than $2,000$ cycles for modular multiplications across various bit-widths. When configured with $8$ CiM macros, LaMoS can complete a $256$-bit modular multiplication in just $32$ cycles. This scaling capability highlights LaMoS's effectiveness in maintaining high performance and efficiency as the number of macros increases.



\section{Conclusion}

In this paper, we propose LaMoS, a scalable SRAM-based CiM architecture for efficient large number modular multiplication. We design a workload segmentation and mapping scheme to map multiplication onto CiM MAC macros and develop LaMoS for parallel acceleration. Additionally, we optimize the mapping to avoid unnecessary macro idles and improve LaMoS's performance at high bit-widths. Extensive experiments demonstrate that LaMoS achieves significant performance advantages compared to existing CiM designs.

\bibliographystyle{IEEEtran}
\bibliography{ref}

@article{xiao2012security,
  title={Security and privacy in cloud computing},
  author={Xiao, Zhifeng and Xiao, Yang},
  journal={IEEE communications surveys \& tutorials},
  volume={15},
  number={2},
  pages={843--859},
  year={2012},
  publisher={IEEE}
}

@article{li2020review,
  title={A review of applications in federated learning},
  author={Li, Li and Fan, Yuxi and Tse, Mike and Lin, Kuo-Yi},
  journal={Computers \& Industrial Engineering},
  volume={149},
  pages={106854},
  year={2020},
  publisher={Elsevier}
}

@article{zhang2021survey,
  title={A survey on federated learning},
  author={Zhang, Chen and Xie, Yu and Bai, Hang and Yu, Bin and Li, Weihong and Gao, Yuan},
  journal={Knowledge-Based Systems},
  volume={216},
  pages={106775},
  year={2021},
  publisher={Elsevier}
}

@inproceedings{liu2023hyperattack,
  title={Hyperattack: An efficient attack framework for hyperdimensional computing},
  author={Liu, Fangxin and Li, Haoming and Chen, Yongbiao and Yang, Tao and Jiang, Li},
  booktitle={2023 60th ACM/IEEE Design Automation Conference (DAC)},
  pages={1--6},
  year={2023},
  organization={IEEE}
}

@inproceedings{li2024hyperfeel,
  title={HyperFeel: An efficient federated learning framework using hyperdimensional computing},
  author={Li, Haomin and Liu, Fangxin and Chen, Yichi and Jiang, Li},
  booktitle={2024 29th Asia and south Pacific design automation conference (ASP-dAC)},
  pages={716--721},
  year={2024},
  organization={IEEE}
}

@article{li2025attack,
  title={Attack and Defense: Enhancing Robustness of Binary Hyper-Dimensional Computing},
  author={Li, Haomin and Liu, Fangxin and Wang, Zongwu and Yang, Ning and Huang, Shiyuan and Liang, Xiaoyao and Guan, Haibing and Jiang, Li},
  journal={ACM Transactions on Architecture and Code Optimization},
  year={2025},
  publisher={ACM New York, NY}
}

@inproceedings{liu2024paap,
  title={Paap-hd: Pim-assisted approximation for efficient hyper-dimensional computing},
  author={Liu, Fangxin and Li, Haomin and Yang, Ning and Chen, Yichi and Wang, Zongwu and Yang, Tao and Jiang, Li},
  booktitle={2024 29th Asia and South Pacific Design Automation Conference (ASP-DAC)},
  pages={46--51},
  year={2024},
  organization={IEEE}
}

@inproceedings{wang2024compass,
  title={Compass: Sram-based computing-in-memory snn accelerator with adaptive spike speculation},
  author={Wang, Zongwu and Liu, Fangxin and Yang, Ning and Huang, Shiyuan and Li, Haomin and Jiang, Li},
  booktitle={2024 57th IEEE/ACM International Symposium on Microarchitecture (MICRO)},
  pages={1090--1106},
  year={2024},
  organization={IEEE}
}

@inproceedings{liu2025asdr,
  title={ASDR: Exploiting Adaptive Sampling and Data Reuse for CIM-based Instant Neural Rendering},
  author={Liu, Fangxin and Li, Haomin and Zhu, Bowen and Wang, Zongwu and Song, Zhuoran and Guan, Haibing and Jiang, Li},
  booktitle={Proceedings of the 30th ACM International Conference on Architectural Support for Programming Languages and Operating Systems, Volume 3},
  pages={18--33},
  year={2025}
}

@inproceedings{liu2025allmod,
  title={ALLMod: Exploring Area-Efficiency of LUT-based Large Number Modular Reduction via Hybrid Workloads},
  author={Liu, Fangxin and Li, Haomin and Wang, Zongwu and Zhang, Bo and Zhang, Mingzhe and Yan, Shoumeng and Jiang, Li and Guan, Haibing},
  booktitle={2025 62nd ACM/IEEE Design Automation Conference (DAC)},
  pages={1--7},
  year={2025},
  organization={IEEE}
}

@article{koblitz1987elliptic,
  title={Elliptic curve cryptosystems},
  author={Koblitz, Neal},
  journal={Mathematics of computation},
  volume={48},
  number={177},
  pages={203--209},
  year={1987}
}

@article{rivest1978method,
  title={A method for obtaining digital signatures and public-key cryptosystems},
  author={Rivest, Ronald L and Shamir, Adi and Adleman, Leonard},
  journal={Communications of the ACM},
  volume={21},
  number={2},
  pages={120--126},
  year={1978},
  publisher={ACM New York, NY, USA}
}

@article{devlin2019blockchain,
  title={Blockchain acceleration using fpgas—elliptic curves, zk-snarks, and vdfs},
  author={Devlin, Ben},
  journal={ZCASH Foundation},
  year={2019}
}

@article{ozturk2020design,
  title={Design and implementation of a low-latency modular multiplication algorithm},
  author={{\"O}zt{\"u}rk, Erdin{\c{c}}},
  journal={IEEE Transactions on Circuits and Systems I: Regular Papers},
  volume={67},
  number={6},
  pages={1902--1911},
  year={2020},
  publisher={IEEE}
}

@inproceedings{gribok2024fpga,
  title={FPGA Modular Multipliers using Hybrid Reduction Techniques},
  author={Gribok, Sergey and Langhammer, Martin and Pasca, Bogdan},
  booktitle={34th International Conference on Field-Programmable Logic and Applications-FPL 2024},
  year={2024}
}

@incollection{mutlu2022modern,
  title={A modern primer on processing in memory},
  author={Mutlu, Onur and Ghose, Saugata and G{\'o}mez-Luna, Juan and Ausavarungnirun, Rachata},
  booktitle={Emerging Computing: From Devices to Systems: Looking Beyond Moore and Von Neumann},
  pages={171--243},
  year={2022},
  publisher={Springer}
}

@inproceedings{barrett1986implementing,
  title={Implementing the Rivest Shamir and Adleman public key encryption algorithm on a standard digital signal processor},
  author={Barrett, Paul},
  booktitle={Conference on the Theory and Application of Cryptographic Techniques},
  pages={311--323},
  year={1986},
  organization={Springer}
}

@article{montgomery1985modular,
  title={Modular multiplication without trial division},
  author={Montgomery, Peter L},
  journal={Mathematics of computation},
  volume={44},
  number={170},
  pages={519--521},
  year={1985}
}

@article{bernstein2017post,
  title={Post-quantum cryptography},
  author={Bernstein, Daniel J and Lange, Tanja},
  journal={Nature},
  volume={549},
  number={7671},
  pages={188--194},
  year={2017},
  publisher={Nature Publishing Group UK London}
}

@article{reis2022imcrypto,
  title={IMCRYPTO: an in-memory computing fabric for AES encryption and decryption},
  author={Reis, Dayane and Geng, Haoran and Niemier, Michael and Hu, Xiaobo Sharon},
  journal={IEEE Transactions on Very Large Scale Integration (VLSI) Systems},
  volume={30},
  number={5},
  pages={553--565},
  year={2022},
  publisher={IEEE}
}

@article{sebastian2020memory,
  title={Memory devices and applications for in-memory computing},
  author={Sebastian, Abu and Le Gallo, Manuel and Khaddam-Aljameh, Riduan and Eleftheriou, Evangelos},
  journal={Nature nanotechnology},
  volume={15},
  number={7},
  pages={529--544},
  year={2020},
  publisher={Nature Publishing Group UK London}
}

@inproceedings{opasatian2023lookup,
  title={Lookup table modular reduction: A low-latency modular reduction for fast ecc processor},
  author={Opasatian, Anawin and Ikeda, Makoto},
  booktitle={2023 IEEE Symposium in Low-Power and High-Speed Chips (COOL CHIPS)},
  pages={1--6},
  year={2023},
  organization={IEEE}
}

@inproceedings{langhammer2021efficient,
  title={Efficient FPGA modular multiplication implementation},
  author={Langhammer, Martin and Pasca, Bogdan},
  booktitle={The 2021 ACM/SIGDA International Symposium on Field-Programmable Gate Arrays},
  pages={217--223},
  year={2021}
}

@book{schneier2007applied,
  title={Applied cryptography: protocols, algorithms, and source code in C},
  author={Schneier, Bruce},
  year={2007},
  publisher={john wiley \& sons}
}

@article{bos2009security,
  title={On the Security of 1024-bit RSA and 160-bit Elliptic Curve Cryptography},
  author={Bos, Joppe and Kaihara, Marcelo and Kleinjung, Thorsten and Lenstra, Arjen K and Montgomery, Peter L},
  year={2009}
}

@article{chen2023digital,
  title={Digital signature standard (DSS)},
  author={Chen, Lily and Moody, Dustin and Regenscheid, Andrew and Robinson, Angela},
  year={2023},
  publisher={Lily Chen, Dustin Moody, Andrew Regenscheid, Angela Robinson}
}

@incollection{goldwasser2019knowledge,
  title={The knowledge complexity of interactive proof-systems},
  author={Goldwasser, Shafi and Micali, Silvio and Rackoff, Chales},
  booktitle={Providing sound foundations for cryptography: On the work of shafi goldwasser and silvio micali},
  pages={203--225},
  year={2019}
}

@inproceedings{fiege1987zero,
  title={Zero knowledge proofs of identity},
  author={Fiege, Uriel and Fiat, Amos and Shamir, Adi},
  booktitle={Proceedings of the nineteenth annual ACM symposium on Theory of computing},
  pages={210--217},
  year={1987}
}

@article{acar2018survey,
  title={A survey on homomorphic encryption schemes: Theory and implementation},
  author={Acar, Abbas and Aksu, Hidayet and Uluagac, A Selcuk and Conti, Mauro},
  journal={ACM Computing Surveys (Csur)},
  volume={51},
  number={4},
  pages={1--35},
  year={2018},
  publisher={ACM New York, NY, USA}
}

@inproceedings{ku2024modsram,
  title={ModSRAM: Algorithm-Hardware Co-Design for Large Number Modular Multiplication in SRAM},
  author={Ku, Jonathan and Zhang, Junyao and Shan, Haoxuan and Samudrala, Saichand and Wu, Jiawen and Zheng, Qilin and Li, Ziru and Rajendran, JV and Chen, Yiran},
  booktitle={2024 61th ACM/IEEE Design Automation Conference (DAC)},
  year={2024}
}

@inproceedings{zhang2021pipezk,
  title={Pipezk: Accelerating zero-knowledge proof with a pipelined architecture},
  author={Zhang, Ye and Wang, Shuo and Zhang, Xian and Dong, Jiangbin and Mao, Xingzhong and Long, Fan and Wang, Cong and Zhou, Dong and Gao, Mingyu and Sun, Guangyu},
  booktitle={2021 ACM/IEEE 48th Annual International Symposium on Computer Architecture (ISCA)},
  pages={416--428},
  year={2021},
  organization={IEEE}
}

@article{li2022mentt,
  title={MeNTT: A compact and efficient processing-in-memory number theoretic transform (NTT) accelerator},
  author={Li, Dai and Pakala, Akhil and Yang, Kaiyuan},
  journal={IEEE Transactions on Very Large Scale Integration (VLSI) Systems},
  volume={30},
  number={5},
  pages={579--588},
  year={2022},
  publisher={IEEE}
}

@inproceedings{zhang2023bp,
  title={Bp-ntt: Fast and compact in-sram number theoretic transform with bit-parallel modular multiplication},
  author={Zhang, Jingyao and Imani, Mohsen and Sadredini, Elaheh},
  booktitle={2023 60th ACM/IEEE Design Automation Conference (DAC)},
  pages={1--6},
  year={2023},
  organization={IEEE}
}

@inproceedings{zhu2022comb,
  title={COMB-MCM: Computing-on-memory-boundary NN processor with bipolar bitwise sparsity optimization for scalable multi-chiplet-module edge machine learning},
  author={Zhu, Haozhe and Jiao, Bo and Zhang, Jinshan and Jia, Xinru and Wang, Yunzhengmao and Guan, Tianchan and Wang, Shengcheng and Niu, Dimin and Zheng, Hongzhong and Chen, Chixiao and others},
  booktitle={2022 IEEE International Solid-State Circuits Conference (ISSCC)},
  volume={65},
  pages={1--3},
  year={2022},
  organization={IEEE}
}

@misc{2019DC,
    author = "{Synopsys}",
    howpublished = {\url{https://www.synopsys.com/community/university-program/teaching-resources.html}},
    year = {[Online]}
}

@article{park2022rm,
  title={Rm-ntt: An rram-based compute-in-memory number theoretic transform accelerator},
  author={Park, Yongmo and Wang, Ziyu and Yoo, Sangmin and Lu, Wei D},
  journal={IEEE Journal on Exploratory Solid-State Computational Devices and Circuits},
  volume={8},
  number={2},
  pages={93--101},
  year={2022},
  publisher={IEEE}
}

@inproceedings{nejatollahi2020cryptopim,
  title={Cryptopim: In-memory acceleration for lattice-based cryptographic hardware},
  author={Nejatollahi, Hamid and Gupta, Saransh and Imani, Mohsen and Rosing, Tajana Simunic and Cammarota, Rosario and Dutt, Nikil},
  booktitle={2020 57th ACM/IEEE Design Automation Conference (DAC)},
  pages={1--6},
  year={2020},
  organization={IEEE}
}

@inproceedings{li2023accelerating,
  title={Accelerating polynomial modular multiplication with crossbar-based compute-in-memory},
  author={Li, Mengyuan and Geng, Haoran and Niemier, Michael and Hu, Xiaobo Sharon},
  booktitle={2023 IEEE/ACM International Conference on Computer Aided Design (ICCAD)},
  pages={1--9},
  year={2023},
  organization={IEEE}
}

\end{document}